# A generic framework for
# adaptive EEG-based BCI training and operation

Author(s):
*Jelena Mladenović (Inria, Bordeaux Sud-Ouest, jelena.mladenovic@inria.fr)
Jeremie Mattout (Inserm, Lyon, jeremie.mattout@inserm.fr)
Fabien Lotte (Inria, Bordeaux Sud-Ouest, fabien.lotte@inria.fr)

## 1. Introduction

Thanks to the technological advancements, the interest in Brain Computer Interfaces (BCI) has grown immensely during the last decades. BCIs are mainly used to facilitate the interaction between people with different disabilities and their environment (Millán 2010, Perrin 2012). However, there have been outreach in non-clinical domains such as gaming and art (Tan and Nijholt 2010, Lotte 2013a).

There are two main paradigms in BCI, depending on the type of extracted physiological markers. **(1)** Spontaneous BCIs (synchronous or self-paced) paradigms, typically measure oscillatory EEG activity, and the event related desynchronisation/synchronisation (ERD/ERS) in Sensorimotor Rhythms (SMR) (Pfurtscheller 2006). They are mainly related to Motor Imagery (MI) BCI, for instance imagining left or right hand movements (Pfurtscheller 2006), and to Mental Imagery, such as mental object rotation or calculations (Faradji 2009). Spontaneous BCI paradigms may also rely on Slow Cortical Potentials (SCP) (Birbaumer and Kübler 2000). **(2)** Evoked responses or ERPs (Event-related Potentials) based BCI paradigms are based on the attentional selection of an external stimulus among many. Be it in the visual (V), the auditory (A) or the somatosensory (S) modality, this approach can give rise to various types of well-known responses such as the P300 component (Donchin 2000) or Steady State sensory evoked potentials, (SS(V/A/S)EP) (Middendorf 2000). Those BCIs follow the same rational, they typically consist of (i) a calibration phase, in which the classifier "learns" to discriminate and translate signal features into commands, (ii) a training phase, in which the user learns to manipulate the system and to regulate his/her EEG patterns, and (iii) the application, in which the user has hopefully control over the system. A general opinion is that there is very little need for user training with ERP-based BCIs (Fazel-Rezai 2012), even though the user can improve his/her P300 marker with training (Baykara 2016). However, the system calibration is often mandatory (Fazel-Rezai 2012) and lasts about 10 minutes. Furthermore, for MI-BCIs, user training is a necessary and often cumbersome process, during which novel functional circuits for action are created, referred to as the "neuroprosthetic skill" (Shenoy 2014, Orsborn 2014). Also, SMR with higher signal-to-noise ratio have been observed as a consequence of learning during such training (Kober 2013; Gaume 2016).

There are two main approaches engaged in improving BCI systems: (i) improving the machine learning techniques (Müller 2008; Makeig 2012), and the newly introduced (ii) improving human learning, by using the knowledge from instructional design and positive psychology (Lotte 2013b; Lotte and Jeunet 2015). Both agree that the system needs to be adapted to the user but rely on different sources of adaptation: the machine for the former and the brain for the latter. In particular, machine learning algorithms should adapt to non-stationary brain signals, while human learning approaches assist in the production of stable EEG patterns of the user, or in the adaptability of the brain to the machine. This implies that these approaches should guide the machine adaptation according to the various users' skills and profiles. Including both aspects of adaptation, a symbiotic co-adaptation (Sanchez 2009) could give rise to a system ready to be used in real life conditions.

However, a major obstacle lies in the large spectrum of sources of variability during BCI use, ranging from (i) imperfect recording conditions: environmental noise, humidity, static electricity etc. (Maby 2016) to (ii) the fluctuations in the user's psychophysiological states, due to: fatigue, motivation or attention etc. (Jeunet 2016). For these reasons a BCI has not yet proved to be reliable enough to be used outside the laboratory (Wolpaw and Wolpaw 2012). Particularly, it is still almost impossible to create one BCI design effective for every user, due to large inter subject variability (Allison & Neuper 2010). Therefore, the main concerns are to create a more robust system with the same high level of success for everyone, at all times, and to improve the current usability of the system (Wolpaw and Wolpaw 2012, Lotte 2013b). This calls for adaptive BCI training and operation.

To our knowledge, there is no work devoted to classifying the literature on adaptive BCI in a comprehensive and structured way. Hence, we propose a conceptual framework which encompasses most important approaches to fit them in such a way that a reader can clearly visualize which elements can be adapted and for what reason. In the interest of having a clear review of the existing adaptive BCIs, this framework considers adaptation approaches for both the user and the machine, i.e., referring to instructional design observations as well as the usual machine learning techniques. It provides not only a coherent review of the extensive literature but also enables the reader to perceive gaps and flaws in current BCI systems, which would, hopefully, bring novel solutions for an overall improvement. BCIs which use non-invasive electroencephalography (EEG) as a measuring tool, will be in the center of our attention throughout this chapter. Nevertheless, the proposed solutions for adaptation can be applied to other techniques such as invasive recordings, functional Near-Infrared Spectroscopy (fNIRS) or Magnetoencephalography (MEG).

This chapter is organised as follows. Section 2 contains a reasoning behind creating adaptive BCIs, it will guide the reader through the aspects of human and machine learning that call for adaptive methods. Section 3 presents our contribution to the field, a comprehensive framework to design and study adaptive BCI systems. We show that the framework encompasses most techniques of adaptive BCIs, which we briefly review. In Section 4 we describe the challenges and future work. Finally Section 5 is the concluding section of this chapter.

## 2. Adaptive BCI systems - motivations and principles

For the sake of understanding the reasons for adaptation, we develop prominent variabilities that cause low BCI performance, and present the main principles used for adaptation.

### 2.1. Reasons for adaptation

Currently, adaptation is mainly done by using different signal processing techniques without including the human factors (Allison and Neuper 2010; Makeig 2012). However, the user's success in mastering the BCI skill appears to be a key element for BCI robustness. If the user is not able to produce stable and distinct EEG patterns, then no signal processing algorithm would be able to recognize them (Lotte 2013b).

Up to a certain extent, machine learning techniques can adapt to the signal variability. However, most of those techniques are blind to the causes of signal variability. Identifying those causes, accounting for them and possibly acting directly on them may help designing more advanced and more robust approaches. Such causes may act at different time scales, for instance a person's drop of attention may have a sudden and dramatic impact, while learning rather operates on the long run.

The term variability is somewhat used to describe the user, environment and equipment "variability", and more frequently, the signal variability. These two variabilities are often confounded, as one being the cause and the other the effect, respectively. Throughout this chapter, we mostly address the user variability as the main cause and denote its various expressions as *components*.

*Causes of signal variability* Variability can be distinguished as: (1) short-term (Schlögl 2010), i.e. signal variabilities within trials or runs caused by, e.g., fluctuations in attention, mood, muscle tension (Jeunet 2016, Schumacher 2015); (2) long-term (Schlögl 2010), such as regulations of sensorimotor rhythms (SMR) over sessions because of learning (Kober 2013). EEG variability can be provoked by many causes, as follows:

The equipment and experimental context:

1. Equipment sensitivity or magnetic field present in the environment (Niedermeyer & da Silva, 2005, Maby 2016);
2. Quality of the instructions given to the user to follow through the task (Neuper 2005).

Short term user *components*:

1. Attention, mood (Nijboer 2008, Jeunet 2016), muscle tension (Schumacher 2015), naturally evolving during, and somewhat driven by, the interaction with a BCI system**.**

2. In the case of no specific instruction, user's mental command itself can be a cause of signal variability. For instance, during a MI task, the user may be using kinesthetic or visual motor imagery as strategy for mental commands (Neuper 2005).

Long term user *components*:

1. The user's learning capacity to control the machine depending on: memory span, intrinsic motivation, curiosity, user profiles and skills  etc. (Jeunet 2016).

Negative or positive loop in learning progression could occur (see instructional design - Keller 2010; Oudeyer 2016). For instance, a positive loop concerns a motivated user whom being motivated has a higher attention level, which would in turn, ideally, enhance learning and control, and finally, induce higher motivation, and so complete the (virtuous) cycle (Mattout 2014).

### 2.2. Main principles of adaptation

When considering adaptation, we mean adaptation of the machine to reduce the negative effect of some user's fluctuations onto the measured signals. In practice, (i) reducing the impact of **signal variability** would require the use of advanced machine learning techniques, such as adaptive spatial filters (Woehrle 2015); (ii) influencing the **user variability** would require adapting the machine output (feedback and instructions) in order to keep the user in an optimal psychological state. The latter could follow instructional design theories by simplifying the layout or diminishing the task difficulty if the user is in a state of fatigue (Sweller 1998; Hattie 2007). Ideally, the BCI system should be (i) set *a priori* for each subject, for instance based on their stable characteristics (skills or profile) and also (ii) dynamically readjusted during the usage, according to their evolving cognitive and affective states.

*Machine Learning* The BCI community has long been aware of the need for adaptive signal processing and classification (Schlögl 2010; Krusienski 2011). Experimental results have confirmed that using adaptive features and classifiers significantly improves BCI performances, both offline and online (McFarland 2011, Mattout. 2014). Signal processing adaptation appears to be particularly useful for spontaneous BCI such as motor imagery (McFarland 2011). However, they can also be useful to reduce calibration time in ERP-based BCI by starting with generic, subject-independent classifiers, and then adapting them to each user during BCI use (Kindermans 2014).

*Human learning* It is shown that one's capacity to create distinct EEG patterns depends amongst others, on one's psychological *components* as: motivation, mood, skills, personality traits etc. (Hammer 2012, Jeunet 2016). In that way, those patterns are more or less detectable and as such influence the BCI performance accuracy (Wolpaw 2002). To assist the users to produce clear EEG patterns is to assist in their learning. To do that, we consider adapting the BCI output (feedback and instructions) by considering a spectre of users' psychological *components* in order to keep them motivated, to perform well, to be efficient and effective. As in any discipline, a well- designed and adapted feedback on one's progress from a tutor is what enables further development, intrinsic motivation and learning (derived from cognitive developmental theories with (Vygotsky 1978), and refined through instructional design theories (Keller 2010)). It is important to design a feedback which would encourage motivation and learning (Lumsden 2016) thus good BCI performance. Moreover, if inappropriate feedback is provided, subjects can learn incorrectly or have negative emotional reactions, which could impair performance and discourage further skill development (Barbero and Grosse-Wentrup 2010).

In order to minimize fatigue and induce motivation in BCI, we should investigate the instructional design theories (Lotte 2013b). These theories could be useful for finding and guiding the adaptation of tasks to users skills, profiles and cognitive abilities. Indeed, it has been shown that, for instance, visual-motor coordination and the ability to concentrate (Hammer 2012), age, gender, practicing sports, gaming, or playing a musical instrument (Randolph 2012), moods and motivation (Nijboer 2008), or spatial abilities (abilities to create, manipulate and transform mental images) (Jeunet 2016) were all positively correlated to motor imagery BCI performances (classification accuracy). It is likely that some other factors may impact BCI performances, e.g. mental workload, which is known to impact learning in general (Sweller 1998). Accounting for the variety of users' psychological *components* would lead to a better BCI feedback design and task adaptation, further assisting them in learning to control a BCI. This way, the EEG patterns would be regulated, which implies that to assist in the user's learning also means to assist in the machine learning techniques.

## 3. Framework

We introduce a conceptual framework which can be used as a tool for a clear visualisation of the elements being adapted, as well as of the missing methods which could possibly lead to optimal adaptive BCI design. It emphasizes existing solutions encompassing all the information possibly used for creating a fully adaptive BCI system. The framework (see Figure 1) has a hierarchical structure, from the lowest level elements which endure rapid changes, to the highest level elements which change at a much slower rate.

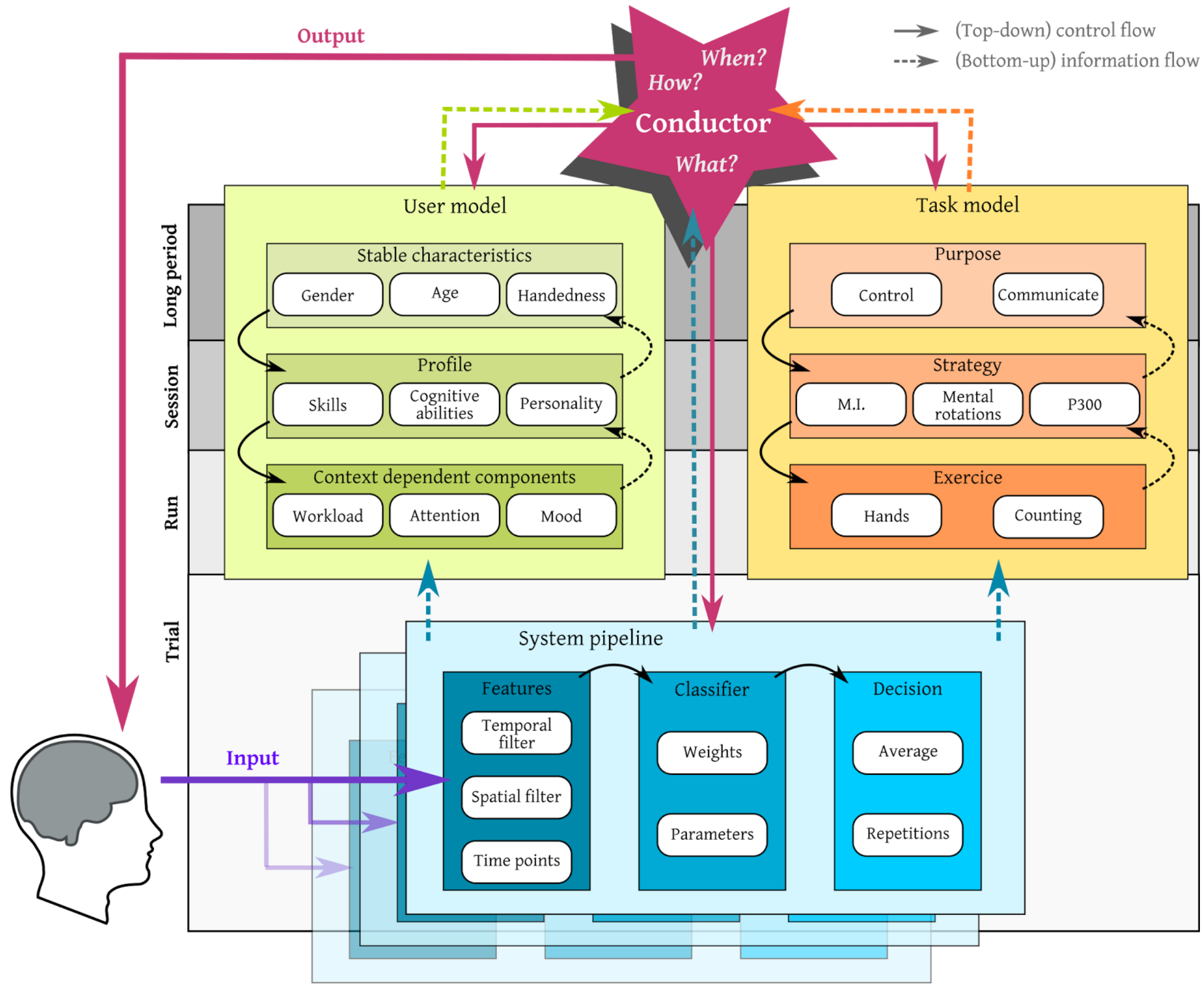

**Figure 1**: Multiple signals (input) maybe observed and processed in parallel in order to infer complementary states or intents, at the trailwise time scale. All the information extracted from these parallel pipelines may trigger the up-dating of the user or task model, which in turn might yield a decision from the conductor to take action, such as adapting one of the systems or the output, or modifying the task or the user model.

It is comprised of 4 major elements, presented bottom-up: **(1)** the **system pipeline** presents the path which the raw EEG signal goes through when manipulated by the computer; **(2a)** the **user model** is an abstraction of the user's states, skills and stable characteristics; **(2b)** the **task model** is a detailed representation of the BCI task; **(3)** the **conductor** masters the adaptation process by deciding the moment, the manner and the elements of the whole system (pipeline, task, user, output) to adapt. The input of the system pipeline comes from (neuro)physiological activity patterns measured on the user, while the output of the system (feedback and instructions) is handled by the conductor and employed by the user. As they undergo rapid changes, input and output take place in the bottom level, as summarized in Figure 1.

To our knowledge, for the first time, we conceptualize a possibility of having an intelligent agent which could eventually replace the experimenter. For the sake of readability, we introduce step by step each element of the framework, starting bottom-up.

### 3.1. The system pipeline

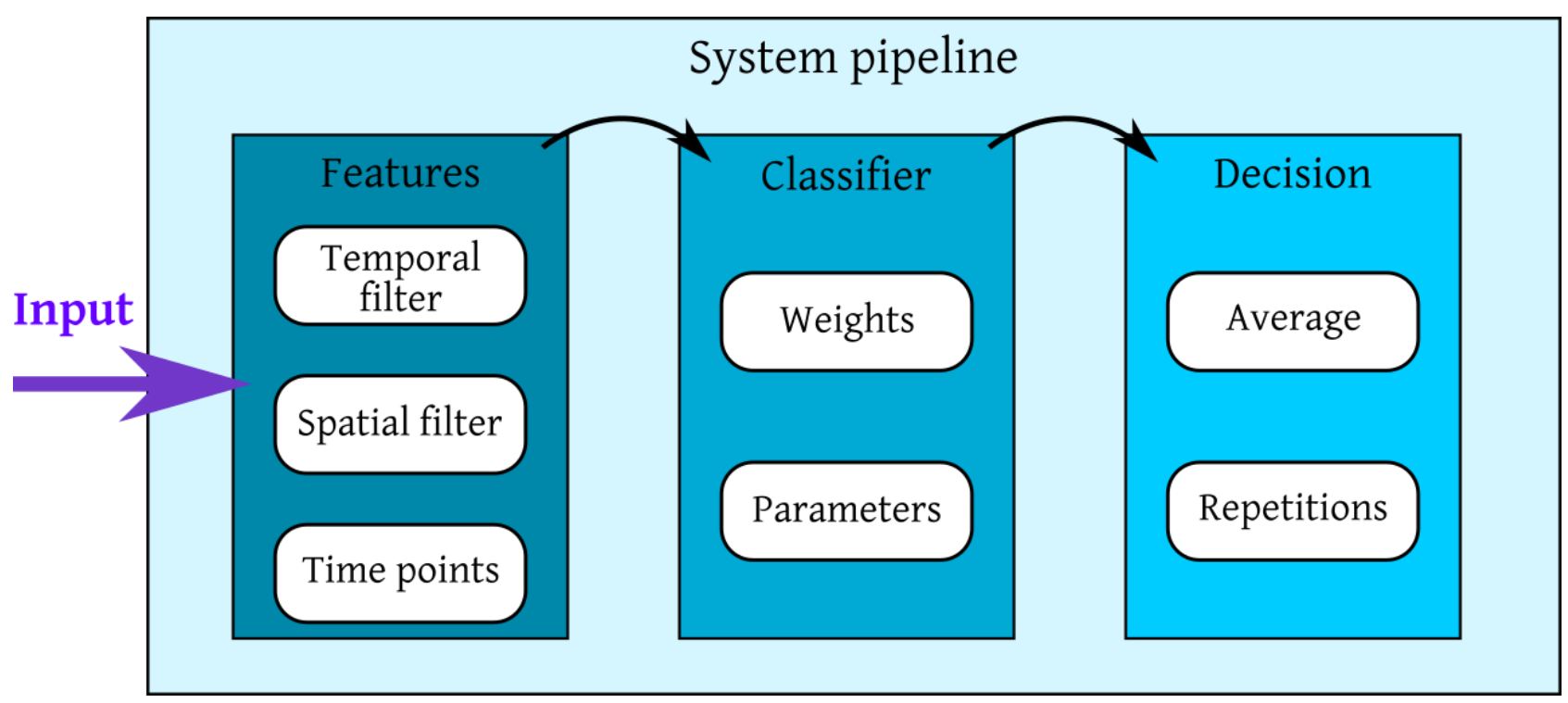

**Figure 2.** The system pipeline, acquiring and processing one of the input signals measured with e.g. EEG

The system pipeline includes: **(1) EEG features** extracted from the raw signal (the input), possibly passing through:

- a temporal filter: to filter noise or to choose an optimal frequency band for instance;
- a spatial filter: combining those electrodes which lead to more discriminating signals; such as Common Spatial Pattern (CSP) filter and its variants.
- signal epoching: selecting a time window to target an event of interest (a motor command or a stimulation);

The extracted EEG features are sent to **(2) the classifier** which translates signal features into the estimated mental commands, using different machine learning classification methods, such as Linear Discriminant Analysis (LDA), whose parameters (e.g. weights) could be adapted.

The accumulation of classification labels over several time samples or epochs give rise to **(3) a decision,** which often defines a speed-accuracy trade-off**.** Typically, with ERP-based systems such as a P300-speller, this is done by accumulating evidence over multiple stimulus repetition, to select a given letter when its probability of being the target letter is higher than a given threshold (Kindermans 2014, Mattout 2014).

In order to maintain or improve BCI performances, one requires to accommodate the signal variability, by adapting either one or several elements of the pipeline:

- Feature extraction, in order to adapt to fast (e.g. a sudden faulty sensor) or slow (e.g. change in the frequency of the signal of interest) changes;
- Classification, in order to change the number of classes, or to change the mapping between each class label and signal features;
- Decision, in order to optimize performance, by adjusting the speed-accuracy trade-off for instance.

## Literature review on adaptive signal processing / machine learning

### Adaptive feature extraction

In order to extract features that adapt to the signal variability, a number of adaptive filters have been proposed for BCI. To the best of our knowledge, they are all supervised, i.e., they require the actual EEG class label. Most of the proposed adaptive filters were spatial filters, and in particular adaptive CSP for motor imagery-based BCI applications. For instance (Sun 2006) and (Shenoy 2006) proposed to re-optimize the CSP filters as a new batch of labelled data becomes available. Later, (Zhao 2008), and (Song 2013) proposed new algorithms to incrementally update the CSP spatial filters without the need to reoptimize everything. (Tomioka 2006) proposed a method to adapt spatial filters to changing EEG data class distribution. Finally, an incrementally adaptive version of the xDAWN spatial filter was proposed (Woehrle 2015), dedicated to ERP based BCI.

Adaptive temporal filters were proposed in (Thomas 2011). In this work, the optimal frequency bands for discriminating motor imagery tasks were regularly re-estimated, and the temporal filters adapted accordingly. It is worth noting that all these adaptive filters algorithms were evaluated only in offline experiments. So, it is unknown how changing the filters influences the users.

Features extracted from EEG signals can be also computed adaptively (Vidaurre 2008). In particular, there are a couple of methods used to estimate features adaptively, with each new EEG sample measured, rather than estimating them as the average feature from a full window of samples in a fixed way. For instance, Adaptive AutoRegressive (AAR) features, estimates AR parameters and use them as features for each new EEG sample (Schlögl 2010), which was proven superior to (fixed) AR parameters estimated on a full time window of samples, including for online experiments. Another example of adaptive features is Adaptive Gaussian Representation, which uses as features time-frequency weights that are adaptively estimated for each time window. (Costa 2000).

Finally, compensating for the features change is possible through the estimation of this change before being used as the classifier input. As such, the corrected features will follow, more or less, the same distribution over time, and thus a classifier trained on features at *t-1*, will still be relevant to classify features at time *t+1*. For instance (Satti 2010) proposed a "Covariate Shift minimization" which first estimates a polynomial function, modeling the moving of the features' distribution center within time. Then, they subtracted this function value at time *t* from the features at the same *t*, to correct for the deviation due to time, which led to improvement of the classification accuracy.

**Adaptive classifiers**

The majority of the work on adaptive signal processing for BCI was so far on the design of adaptive classifiers, i.e., classifiers whose parameters were incrementally re-estimated over time. Both supervised and unsupervised (not having the class labels) adaptive classifiers were proposed.

In the supervised category, multiple classifiers were explored offline including Gaussian classifiers (Buttfield 2006), LDA or Quadratic Discriminant Analysis (QDA) (Shenoy 2006, Schlögl 2010) for mental imagery-based BCI. For ERP-based BCI, (Woehrle 2015) explored adaptive Support Vector Machine (SVM), adaptive LDA, a stochastic gradient-based adaptive linear classifier, and Online passive-aggressive algorithms. Online, still in a supervised way, only the LDA/QDA (Vidaurre 2007) and an adaptive variational bayesian classifier (Sykacek 2004) were explored.

Unsupervised adaptation of the classifiers is obviously much more difficult since the class labels, hence the class specific variability, is unknown. Thus, unsupervised methods were proposed that try to estimate the class labels of the new incoming samples first, before adapting the classifier based on this estimation. This was explored offline in (Blumberg 2007) and (Gan 2006) for an LDA classifier with motor imagery data. Another simple unsupervised adaptation of the LDA classifier for motor imagery data was proposed and evaluated both offline and online in (Vidaurre 2011a). The idea is not to incrementally adapt all the LDA parameters, but only its bias, which can be estimated without knowing the class labels if we know that the data are balanced, i.e., with the same number of samples per class.

For ERP-based BCI, semi-supervised learning also proved useful for adaptation. Semi-supervised learning consists in using a supervised classifier to estimate the labels of unlabelled data, so to adapt this classifier, based on these initially unlabelled data. This was explored with SVM and enabled to calibrate P300-spellers with less data than with a fixed, non-adaptive classifier (Li 2008, Lu 2009).

Finally, both offline and online, (Kindermans 2014) proposed a probabilistic method to adaptively estimate the parameters of a linear classifier in P300-based spellers, which led to drastic reduction in calibration time, essentially removing the need for the initial calibration altogether. This method exploited the specific structure of the P300-speller, and notably the frequency of samples from each class at each time, to probabilistically estimate the most likely class label. In a related work, (Grizou 2014) proposed a generic method to adaptively estimate the parameters of the classifier without knowing the true class labels by exploiting any structure that the application may have.

### Fully adaptive signal processing

It is possible to use fully adaptive BCI signal processing pipelines. Several groups have explored BCI designs with both adaptive features and classifiers.

Offline adaptive xDAWN and several adaptive classifiers for ERP-based BCIs are studied in (Woehrle 2015), showing that each improved performances as compared to a non-adaptive version. Even so, combining them both improved the classification accuracy even further.

Online for motor imagery based BCI, (Vidaurre 2007) explored using both Adaptive AR features with an Adaptive LDA. Later, she also explored co-adaptive training, where both the machine and the user are continuously learning, by using adaptive features and an adaptive LDA classifier (Vidaurre 2011b). This enabled some users, who were initially unable to control the BCI, to reach classification performances better than by chance. This work was later refined in (Faller 2012) by using a simpler but fully adaptive setup with auto-calibration, which proved to be efficient including for users with disabilities (Faller 2014). Co-adaptive training, using adaptive CSP Patches proved to be even more efficient (Sannelli 2016).

Altogether, these studies clearly stress the benefits of adaptive signal processing for EEG-based BCI, both at the feature extraction, classifier and decision levels. However, these works often omit the human factors.

### Adaptive decision methods

The decision can be adaptive as well, by adapting the speed-accuracy trade-off for wheelchair control (Saedi 2015), or adapting the number of repetitions in a P300 speller (Mattout 2014). While monitoring the user's state, it is also possible to inhibit BCI interaction until specific requirements, such as the user attention level are met (George 2011).

Monitoring the user during the BCI task could be useful for revealing a way to adapt features over time, such as an increase in workload which can impact MI features (Gerjets 2015). Hence, we introduce the user, and the task they have to do, as a guide for adapting the system.

### 3.2. The User and the Task Model

In order to adapt all the elements of the system pipeline with respect to the user skills and states, it is useful to consider a user model (Figure 3). We assume that the user *components* have a degree of changeability within certain time intervals and also react to the machine output. Hence, we categorized the user model according to time, within a timeline based on 3 time scales: runs, sessions and a loosely long time period. To create a complete automatic adaptation would mean to refine the machine to manage more precisely the user's responses. For that purpose, we created a task model (Figure 4), containing the necessary BCI task information, which components follow the same time intervals as the user model.

The timeline prescribes how often the system should be adapted/updated and according to which element. Notably, the time intervals are chosen as they are commonly used in the BCI community, but it is not necessary to have them fixed as such.

#### User model

We accustomed the Scherrer's classification of affective states (Martín 2013) for the BCI purposes, and arranged them in the user model. Namely, the user model is an abstraction of the user, where their skills, states, and stable characteristics are arranged according to the time needed for them to change. The more we climb up, the more stable the components are.

**Figure 3**. User model, containing 3 levels, arranged from the least stable (context dependent), to the most stable components (stable characteristics).

- *Stable user characteristics*: gender, age, culture, background, genetic predispositions (handedness) etc. These elements can assist in accounting for inter subject variability.
- *User profile:* (i) a given user may have developed particular (non-BCI or BCI) skills which may help in the current BCI context or may be reinforced by the ongoing practice. (ii) same for personality traits (openness, conscientiousness, extraversion, agreeableness, neuroticism, flow proneness etc.); (iii) and cognitive abilities (memory span, imagination, attention span etc.);
- *Context dependent characteristics*: the user's cognitive and affective state (attention level, fear, stress, etc.) are very much related to the current task set and environmental situation.

**A brief literature review of "adaptive methods" related to the user model**

There are many attempts to predict the users' performance (predictors) in order to fully customize the system to users' needs. Hence, by knowing that some of their already acquired skills relate to the BCI skills, such as spatial abilities (Teillet, 2016), they can be trained and improved beforehand, without using a BCI. This way, by improving those skills, one improves a BCI skill as well. Table 1 gathers several predictors and BCI training methods for each element of our user model. For a detailed report on predictors, see (Jeunet 2016).

| | Stable characteristics | Profile | Context dependent components |
|---|---|---|---|
| Predictors | Age determining performance (Zich 2016)<br><br>Paraplegic (Vuckovic 2014)<br><br>Gender (Randolph 2012) | Acquired skill (gaming, sport - Randolph 2012)<br>Visual-motor coordination (Hammer 2012)<br>Spatial abilities (Jeunet 2016)<br><br>High θ and low α powers reveal illiteracy (Ahn 2013) | Mood and motivation, confidence (Nijboer 2008), Attention (Hammer 2012), Fear of BCIs (Witte 2013)<br><br>γ oscillations (Grosse-Wentrup 2012) |
| Training adaptation | / | Spatial ability training (Teillet 2016) | Mindfulness training (Tan 2014)<br><br>Attenuating γ-power for good BCI-performance (attention) (Grosse-Wentrup 2011) |

**Table 1.** Examples of predictors and user training methods regarding each **user model component**.

**TASK model**

The goal of the task model is to assist the BCI user in accomplishing his/her goal (communication/control, rehabilitation, amusement or artistic expression). Similarly to the user model, the task model can be organized hierarchically according to the three following time scales: runs, sessions and long period.

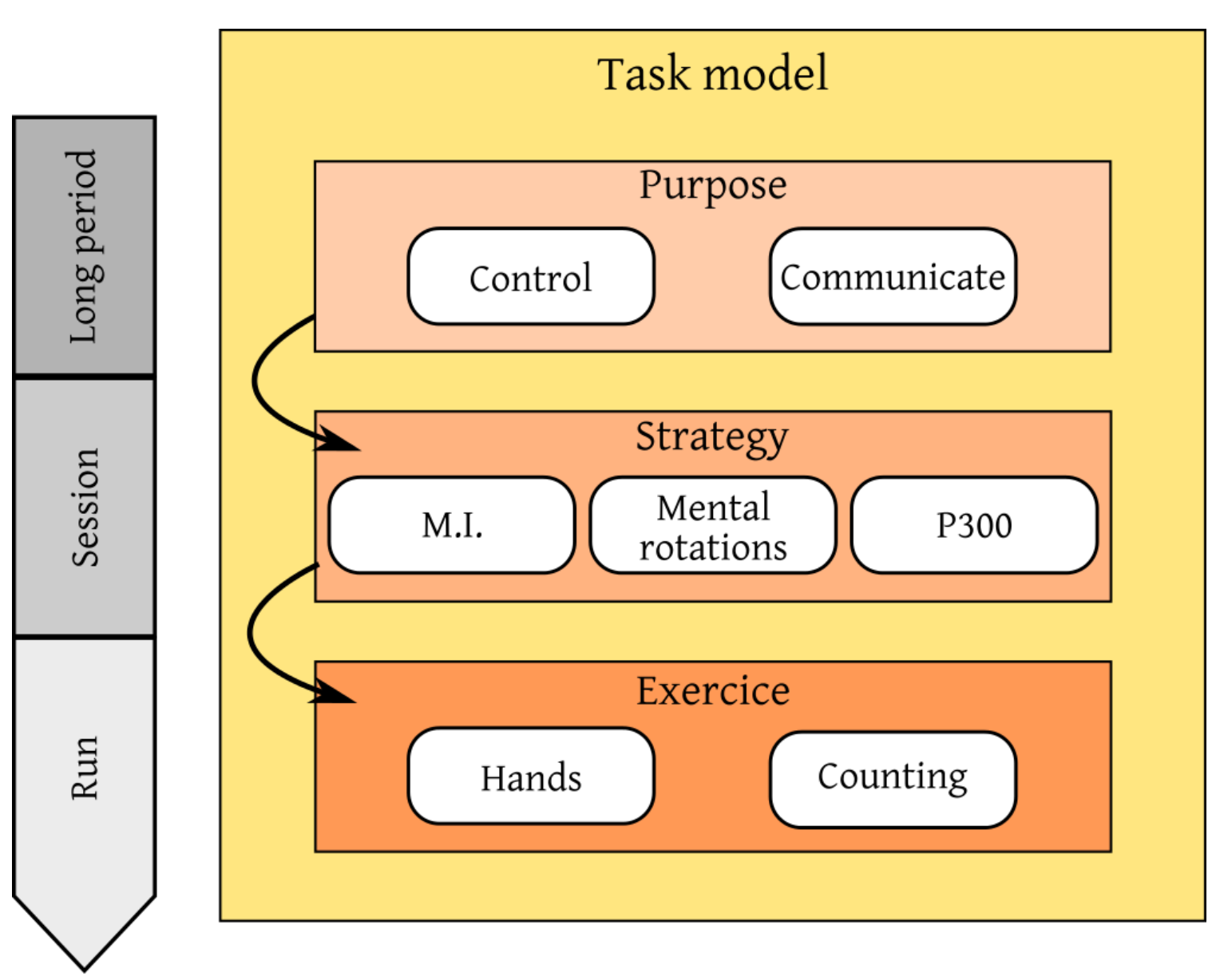

**Figure 4**. Task model, arranged within 3 time scales.

Components of the task model are typically determined beforehand, by the experimenter, and are not changed during the BCI session. Nevertheless, we envision the possibility to adapt each of these elements within the timeline. The task model is comprised of:

- *Purpose of a BCI*: (1) a tool to control: prostheses, wheelchairs and other devices; (2) a communication device: writing words on a screen, etc; (3) a tool for rehabilitation; after stroke (Birbaumer 2007), for paraplegic patients (Vuckovic 2014), for autistic and ADHD children (Friedrich 2014), and others; (4) a tool for artistic expression (creating music or paintings) or entertainment (Lécuyer 2008), etc.
- *Strategy*: the most used strategies include Mental Imagery, P300 or SSEP. The strategy will influence the choice of the initial signal processing and classification techniques, for instance using the Motor Imagery BCI strategy, the bandpower initially considered could be 8-12 Hz (mu rhythm) and 13-30 Hz (beta rhythm), measured on the electrodes placed over the sensorimotor cortex, while P300 would mean considering band-pass filtered time series (between 1-20 Hz) on fronto-central, parietal and occipital regions.
- *Exercise:* It indicates the mental command to be used given a strategy, as for MI strategies, an exercise is chosen between various motor imageries such feet, hands, or tongue movements.

The strategy and exercise, initialized by the BCI purpose, can be adapted automatically based on the evolution of the user's need or state, as informed by bottom-up message passing. For instance, the user's performance being lower than a certain threshold could indicate the need to change strategy or exercise.

**A brief literature review of adaptive methods related to the task model**

*Purpose* Depending on the purpose of the BCI, the adaptation methods will differ. Rehabilitation will favour methods engaged in learning and self-regulation, Communication will favour methods that improve accuracy and speed, while application for entertainment will favour design and innovation etc. We have not found literature fostering this idea, thus it should be left as a perspective for future adaptive BCIs.

*Strategies* A strategy can be switched to another, favouring the one in which the user produces the clearest EEG patterns and has the highest performance. In (Pfurtscheller 2010, Müller-Putz 2015), the use of Hybrid BCIs is suggested, i.e., switching between or using multiple BCI strategies (e.g. P300, MI, SSEP); or combining different measuring techniques (e.g. M/EEG or fNIRS); or using tools apart from those in BCI, such as eye trackers, electrocardiograms, etc. To account for these possible hybrid BCIs is why there are multiple instances of the system pipeline in Figure 1.

*Exercises* An exercise depends on the chosen strategy. Adapting exercises within runs -- such as variating between hand or foot imagination, choosing the one that had better performance (Friedrich. 2013; Fruitet. 2013). Adaptation is possible in larger time intervals such as within sessions, e.g. switching from 1D, 2D, to 3D MI-BCI tasks (McFarland. 2010), changing everything related to it (the instructions and feedback).

### 3.3. Machine Output

Along with feedback, instructions are what the user can receive from the machine. They can be adapted by the conductor based on the information flowing bottom-up, from the various signal processing pipelines or systems, through the user and task models.

#### Feedback

The feedback is usually a representation of the classifier's output, managed by the decision. It can be seen as the machine's response to the user's performance or states. It is useful or even necessary for the user's self-regulation process or learning, to be informed on his/her progress when accomplishing the task. There are many different types of feedback, supporting emotional or cognitive states of the user, possibly given in: (i) **different modalities**, typically visual (Neuper 2005), but also vibrotactile (Jeunet 2015), a tangible interface (Frey 2015), or an immersive virtual environment (Vourvopoulos 2016); and (ii) **degrees of assistance** (biased feedback - Barbero 2010). A feedback could be designed to target some *stable user characteristics*, such as applications for autistic children (Friedrich 2014), or *context dependent* user components, such as inducing motivation in a social context (Bonnet 2013), or catching the user's attention with video games (Angevin 2008).

Adapting feedback could potentially bring benefits which favor ergonomy, minimize fatigue, and optimize learning. This mostly remains to be explored.

**Instructions and stimuli**

These are, for instance the stimuli (flashing letters) in P300-spellers, or arrows indicating the user to perform left or right hand motor imagery in MI BCI. The instructions could be presented/adapted, independently from the classifier's output in: (i) **different modality***:* visual, auditive or tactile; or (ii) **difficulty***:*

- Speed -- the speed of instruction's appearance might decrease over time, if we assume the user's fatigue (decreasing the task difficulty)
- order of appearance -- it would be interesting to investigate whether presenting a block of instructions for one-class motor imagery (arrow left-hand) and then a block of the other class (right hand) is easier for some users than presenting them in an alternate manner (left-right).

**3.4. Conducting adaptation with the conductor**

As each of the framework elements can be adapted/updated separately, or in combination, using various algorithms or criteria, we explicitly refer to a controlling agent in our framework, which would preferably be created for a global adaptive BCI. It gathers all the information available from the user, the task, and the signal processing pipelines, in order to decide the how, when and what to adapt. The conductor would need an objective function, upon which it would make its decisions. We draw an analogy with Intelligent Tutoring Systems (ITS), which are methods creating objective metrics and computational models for learning with digital environment. With our adaptive framework, our user is the ITS student, the conductor is the ITS tutor, and the task the ITS expert (NKambou 2010).

ITSs adapt content and activities for the purpose of challenging and guiding students in an optimal way, i.e., preventing them from being too overwhelmed with difficult material or too bored with easy or repetitive material (Murray and Arroyo 2002). There are many methods dealing with adapting the content of the task to keep students' attention and motivation up, and most of them are inspired by the following two approaches: (i) maintaining the zone of proximal development (ZPD) (Vygotsky 1978); (ii) being in flow (Nakamura and Csikszentmihalyi 2014). The first, based on cognitive developmental theory for instructional design (Luckin 2001), may guide an indirect estimation of the person's cognitive resources (Allal and Ducrey 2000). Flow, originating from positive psychology, is an autotelic (self-rewarding) state, where ons is immersed in a task so that one loses the sense of time, of self, and of the environment. They both concord with theories of intrinsic motivation which suggest that motivation and learning improve if the proposed exercises are at a level that is slightly higher than the current user's skill level.

Choosing automatically the optimal task, in real time, while considering the user and task models, could bring promising results for BCI training and operation. For instance, the conductor, as for some ITS, could use Multi-Armed Bandit algorithms to select an optimal sequence of tasks and outputs (Clement 2015).

**4. Perspectives and challenges**

The perspectives we consider here correspond to some gaps we noticed while confronting the literature we are aware of, to the proposed new framework. First of all, the gaps in the user model: training methods (outside the BCI context), and feedback and instructions (during the BCI task) should adapt considering user's: *(i) stable characteristics* -- considering patients with different disorders (paraplegic, after stroke, autistic etc), also considering different preferences (between children and adults, women and men etc); *(ii) profile* -- for individuals who differ in their skills or personality traits; *(iii) context dependent characteristics,* favouring those methods which increase attention level, motivation etc.

Another important matter, instead of adapting the nature of the exercise based on the user's performance only (typically the classifier's output), we could also account for context dependent components, such as the user's attention or workload level, as monitored with some passive BCI pipeline or with other physiological sensors (e.g. skin conductance).

The challenges we encounter when considering the full adaptation with the conductor are: (1) identifying metrics and criteria to optimize depending on the task, to ensure relevant adaptation, i.e. favouring those adaptation methods which most concur to user's needs; (2) designing computational models of the user and task models; (3) testing the adaptive BCI online and validating it with real experiments; (4) designing unsupervised adaptive features and classifiers, and validating them online (most of them are supervised and only evaluated offline so far); (5) propose adaptive feedback and exercises.

As for the conductor, beside the algorithm that should decide when and how to perform the adaptations, the criteria for adapting the whole system is hidden in the "purpose" of the BCI, or what one wishes to achieve. Hence, will the conductor aim for flow or ZPD as an objective function (by adapting the task difficulty or by presenting a biased feedback for example) or for system's performance and speed (by favouring higher classification accuracy). Finally, we need to ensure that adapting the system will not impede the inevitable user adaptation (human learning), and thus lead to a virtuous co-adaptation.

## 5. Conclusion

Throughout this chapter, we emphasized the crucial need for adaptive methods in order to optimize the design and online performance of BCI. We stressed out the fact that, in order to create an overall adaptive system, it is not sufficient to consider adapting the signal processing and classification techniques, but also the output and the task parameters, in order to fully accommodate the user's variability in terms of needs and psychophysiological states. Following that requirement, we created a framework, comprised of: (i) one or several BCI systems/pipelines; (ii) a user model, whose elements are arranged according to different time scales ; (iii) a task model, enabling the system adaptation with respect to the user model; (iv) the conductor, an intelligent agent which implements the adaptive control of the whole system. For the first time, we conceptualize a fully adaptive BCI system, with respect to the user needs and states. The existing adaptation methods are described through an extensive literature review of each element of both types of models (user and task) and of possible low-level pipelines for raw signal processing.

The potential benefits of using this framework are numerous, for one it enables clear and methodological visualization of all the BCI system components, their possible interaction, the way and the context in which they could be adapted. Although invasive brain activity recording approaches (e.g. ECoG or intracortical electrodes array) consider different signal processing algorithms from those presented in our pipeline, the same principles for co-adaptivity (Sanchez 2009) and the rest of the framework structure and methods apply for any BCI system, be it invasive or not. Moreover, this framework is also convenient for mapping the literature onto each of its components in order to understand current issues in BCI in general, and to visualize the gaps to be filled by future studies in order to further improve BCI usability. We believe this framework will contribute to delve possible future research paths, and give rise to novel challenges and ventures.

## Acknowledgment

This work was supported by the French National Research Agency with the REBEL project and grant ANR-15-CE23-0013-0.

## References

Ahn, M., Cho, H, Ahn, S, Jun, S.C. 2013. *High Theta and Low Alpha Powers May Be Indicative of BCI-Illiteracy in Motor Imagery*. PLoS ONE 8(11): e80886. doi:10.1371/journal.pone.0080886

Allal, L., and Ducrey, G.P. 2000. *Assessment of—or in—the zone of proximal development*. Learn.Instr. 10, 137–152.

Allison, B., and Neuper, C. 2010. *Could Anyone Use a BCI?* Springer London.

Barbero, Á., Grosse-Wentrup, M. 2010. *Biased feedback in brain-computer interfaces*, Journal of NeuroEng and Rehab

Baykara, E., Ruf, C.A., Fioravanti, C., Käthner, I., Simon, N., Kleih, S.C., Kübler, A., Halder, S. 2016. *Effects of training and motivation on auditory P300 brain-computer interface performance*, Clin Neurophy;127(1):379-87. doi: 10.1016/j.clinph.2015.04.054.

Birbaumer, N., and Kübler, A. 2000. *The thought translation device (TTD) for completely paralyzed patients*, IEEE Trans. Rehab. Eng., vol. 8, pp. 190–193

Birbaumer, N., Cohen, L.G. 2007. *Brain-computer interfaces: Communication and restoration of movement in paralysis*. J. Phys. 579, 621–636.

Blumberg, J., Rickert, J. Waldert, S. Schulze-Bonhage, A. Aertsen, A. & Mehring, C. 2007. *Adaptive Classification for Brain Computer Interfaces*, Proc. EMBC, 2536-2539

Buttfield, A., Ferrez, P. & Millan, J. 2006. *Towards a robust BCI: error potentials and online learning.* IEEE Trans on Neur Sys and Rehab Eng, *14*, 164-168

Clement, B., et al. *Multi-Armed Bandits for Intelligent Tutoring Systems*. 2015. JEDM-Journal of Educational Data Mining 7.2: 20-48.

Costa, E. & Jr, E. C. 2000. *EEG-based discrimination between imagination of left and right hand movements using adaptive gaussian representation. Med Eng & Physics, 22*, 345-348

Donchin E, Spencer K. V., and Wijesinghe R. 2000. *The mental prosthesis: Assessing the speed of a P300-based brain–computer interFace*. IEEE Trans. Rehab. Eng., vol. 8, no. 2, pp. 174–179.

Faller, J., Vidaurre, C., Solis-Escalante, T., Neuper, C., & Scherer, R., *Autocalibration and recurrent adaptation: Towards a Plug and Play Online ERD-BCI*. 2012. *IEEE Trans Neural Sys and Rehab Eng, 20*, 313-319

Faller, J., Scherer, R., Costa, U., Opisso, E., Medina, J. & Müller-Putz. 2014. G.R. *A Co-Adaptive Brain-Computer Interface for End Users with Severe Motor Impairment, PloS one, 9*, e101168

Faradji, F., Ward, R.K., Birch, G.E. 2009. *Plausibility assessment of a 2-state self-paced mental task-based BCI using the no-control performance analysis* J. Neurosci. Methods, 180, pp. 330–339

Fazel-Rezai, R., Allison, B., Guger, C., Sellers, E., Kleih, S. & Kübler, A. 2012. *P300 brain computer interface: current challenges and emerging trends*, Frontiers in Neuroeng, 5

Frey, J., Gervais, R., Fleck, S., Lotte, F., & Hachet, M. 2014. *Teegi: tangible EEG interface*. In Proc 27th ACM symposium UIST (pp. 301-308).

Friedrich, E.V., Neuper, C., & Scherer, R. 2013. *Whatever works: a systematic user-centered training protocol to optimize brain-computer interfacing individually*. PloS one, 8(9), e76214.

Friedrich EV, Suttie N, Sivanathan A, Lim T, Louchart S, Pineda JA. 2014. *Brain-computer interface game applications for combined neurofeedback and biofeedback treatment for children on the autism spectrum*. Front Neuroeng.7:21. doi: 10.3389/fneng.2014.00021. eCollection 2014.

Fruitet, J., Carpentier, A., Munos, R., & Clerc, M. 2013. *Automatic motor task selection via a bandit algorithm for a brain-controlled button*. Journal of neuroeng, 10(1), 016012.

Gan J., *Self-adapting bci based on unsupervised learning*. 2006. Proc. Int. BCI workshop

George L, Bonnet L, Lécuyer A. 2011. *Freeze the BCI until the user is ready: a pilot study of a BCI inhibitor*, 5th International BCI Workshop

Gerjets, P., Walter, C., Rosenstiel, W., Bogdan, M., & Zander, T. O. 2015. *Cognitive state monitoring and the design of adaptive instruction in digital environments: lessons learned from cognitive workload assessment using a passive brain-computer interface approach*. Using Neurophysiological Signals that Reflect Cognitive or Affective State, 20.

Grosse-Wentrup. 2011. *Neurofeedback of Fronto-Parietal Gamma-Oscillations.* In 5th International BCI Conf (pp. 172-175).

Grosse-Wentrup & Scholkopf. 2012. *High Gamma-Power Predicts Performance in SMR BCIs*, Journal of NeuroEng

Grizou, J.; Iturrate, I.; Montesano, L.; Oudeyer, P.-Y. & Lopes. 2014. M. *Calibration-free BCI based control* Proc. AAAI,1-8

Gaume, A., Vialatte, A., Mora-Sánchez, A., Ramdani, C., & Vialatte, F. B. 2016. *A psychoengineering paradigm for the neurocognitive mechanisms of biofeedback and neurofeedback*. Neuroscience & Biobehavioral Reviews, 68, 891-910.

Hammer, E. M., Halder, S., Blankertz, B., Sannelli, C., Dickhaus, T., Kleih, S., ... & Kübler, A. 2012. *Psychological predictors of SMR-BCI performance*. Biological psychology, 89(1), 80-86

Hattie, J. & Timperley, H. 2007. *The Power of Feedback Review of Educational Research*, 77, 81-112

Jeunet, C., Vi, C., Spelmezan, D., N'Kaoua, B., Lotte, F., & Subramanian, S. 2015. *Continuous tactile feedback for motor-imagery based brain-computer interaction in a multitasking context*. HCI (pp. 488-505). Springer International Publishing.

Jeunet C., N'Kaoua, B., & Lotte, F. 2016 *Advances in user-training for mental-imagery-based BCI control: Psychological and cognitive factors and their neural correlates*, Progress in brain research

Kindermans, P.J., Tangermann, M., Müller, K.-R. & Schrauwen, B., 2014 *Integrating dynamic stopping, transfer learning and language models in an adaptive zero-training ERP speller,* Journal of neural engineering, 11, 035005

Kober, S.E., Witte, M., Ninaus, M., Neuper, C. and Wood, G., 2013. *Learning to modulate one's own brain activity: the effect of spontaneous mental strategies.*

Krusienski, D.; Grosse-Wentrup, M.; Galán, F.; Coyle, D.; Miller, K.; Forney, E. & Anderson, C. 2011. *Critical issues in state-of-the-art brain–computer interface signal processing*, Journal of NeuoEng, 8, 025002

Keller, J. 2010. *Motivational design for learning and performance: The ARCS model approach* Springer,

Lécuyer, A,, Lotte, F., Reilly, R.B., Leeb, R., Hirose, M., and Slater, M. 2008. *Brain computer interfaces, virtual reality, and videogames*. IEEE Computer, 41(10):66–72.

Li Y.; Guan, C.; Li, H. & Chin, Z. 2008. *A self-training semi-supervised SVM algorithm and its application in an EEG-based brain computer interface speller system*. Pattern Recognition Letters,29, 1285-1294

Lotte, F., Faller, J., Guger, C., Renard, Y., Pfurtscheller, G., Lécuyer, A., Leeb, R., 2013a. *Combining BCI with Virtual Reality: Towards New Applications and Improved BCI. In: Towards Practical Brain-Computer Interfaces*. Pp. 197-220. Springer Berlin Heidelberg.

Lotte, F., Larrue, F., & Mühl, C. 2013b. *Flaws in current human training protocols for spontaneous brain-computer interfaces: lessons learned from instructional design.*

Lotte, F., & Jeunet, C. 2015. *Towards improved bci based on human learning principles. In Brain-Computer Interface*, 3rd International Winter Conference on (pp. 1-4). IEEE.

Lu, S.; Guan, C. & Zhang, H. 2009. *Unsupervised Brain Computer Interface based on Inter-Subject Information and Online Adaptation*, IEEE Transactions on Neural Systems and Rehabilitation Engineering, 17, 135-145

Luckin, R. 2001. *Designing children's software to ensure productive interactivity through collaboration in the zone of proximal development (zpd)*. Info Tech in Childhood Education Annual, 1, 57–85

Lumsden J, Edwards E, Lawrence N, Coyle D and Munafò M, 2016. *Gamification of Cognitive Assessment and Cognitive Training: A Systematic Review of Applications and Efficacy,* JMIR Serious Games vol: 4(2)pp: e11

Maby E., 2016. chapters *7. Sensors: theory and innovation*, *8. Technical requirements for high quality EEG recordings*, in *Brain-Computer Interfaces 2: technology and applications*, Ed. Clerc M, Bougrain L, Lotte F, ISTE-Wiley

Makeig, S., Kothe, C., Mullen, T., Bigdely-Shamlo, N., Zhang, Z., & Kreutz-Delgado, K. 2012. *Evolving signal processing for brain–computer interfaces*. Proceedings of the IEEE, 100, 1567-1584.

Mattout J, Perrin M, Bertrand O, Maby E. 2014. *Improving BCI performance through co-adaptation: applications to the P300-speller;* 58(1):23-8. doi: 10.1016/j.rehab.2014.10.006.

McFarland, D.; Sarnacki, W. & Wolpaw, J. 2010. *Electroencephalographic (EEG) control of three-dimensional movement.* Journal of NeurEng

McFarland, D.; Sarnacki, W. & Wolpaw, J. 2011. *Should the parameters of a BCI translation algorithm be continually adapted?* Journal of Neuroscience Methods, 199, 103 - 107

Müller-Putz, G., Leeb, R., Tangermann, M., Höhne, J., Kübler, A., Cincotti, F., ... & Millán, J. D. R. 2015. *Towards noninvasive hybrid brain–computer interfaces: framework, practice, clinical application, and beyond*. Procs of the IEEE, 103(6), 926-943.

Murray T & Arroyo I. 2002. *Toward Measuring and Maintaining the Zone of Proximal Development in Adaptive Instructional Systems*, Inter Conf on ITS

Martín E , Haya P.A., Carro R.M. 2013. *Modeling and Adaptation for Daily Routines: Providing Assistance to People with Special Needs*, Springer Sci & Bus Media

Middendorf M., McMillan G., Calhoun G., and Jones S. K. 2000. *Brain–computer interfaces based on the steady-state visual-evoked response*. IEEE Trans. Rehab. Eng. , vol. 8, no. 2, pp. 211–214.

Millán J. d R, Rupp R, Müller-Putz G, Murray R., Smith R., Giugliemma C., Tangermann M., Vidaurre C., Cincotti F., Kübler A., Leeb R., Neuper C., Müller K.R., and Mattia D. 2010. *Combining brain–computer interfaces and assistive technologies: stateoftheart and challenges*. Front in neurosci, 4.

Müller, K. R., Tangermann, M., Dornhege, G., Krauledat, M., Curio, G., & Blankertz, B. 2008. *Machine learning for real-time single-trial EEG-analysis: from brain–computer interfacing to mental state monitoring.* J of neurosci methods, 167(1), 82-90.

Nakamura, J., & Csikszentmihalyi, M. 2014. *The concept of flow. In Flow and the foundations of positive psychology* (pp. 239-263). Springer Netherlands.

Neuper C., Scherer R., Reiner M., and Pfurtscheller G. 2005. *Imagery of motor actions: Differential effects of kinesthetic and visual-motor mode of imagery in single-trial EEG*, Brain Res. Cogn. Brain Res., vol. 25, no. 3, pp. 668–677

Niedermeyer, E. & da Silva, F. L. 2005. *Electroencephalography: basic principles, clinical applications, and related fields* Lippincott Williams & Wilkins

Nijboer, F., Furdea, A., Gunst, I., Mellinger, J., McFarland, D. J., Birbaumer, N., & Kübler, A. 2008. *An auditory brain–computer interface (BCI)*. Journal of neuroscience methods, 167(1), 43-50.

NKambou, R., Mizoguchi, R. and Boudreau, J. *Advances in intelligent tutoring systems*. Vol. 308. Springer. 2010.

Orsborn, A.L., Moorman, H.G., Overduin, S.A., Shanechi, M.M., Dimitrov, D.F. and Carmena, J.M., 2014. Closed-loop decoder adaptation shapes neural plasticity for skillful neuroprosthetic control. *Neuron*, *82*(6), pp.1380-1393.

Oudeyer P-Y, Gottlieb, J, Lopes, M. 2016. *Intrinsic motivation, curiosity, and learning: Theory and applications in educational technologies*:. Prog in brain res,

Perrin, M.; Maby, E.; Daligault, S.; Bertrand, O.; Mattout, J. 2012. *Objective and subjective evaluation of online error correction during P300-based spelling*. Adv. HCI 2012, 578295:1–578295:13.

Pfurtscheller, G., Brunner, C., Schlögl, A., Lopes da Silva F.H. 2006. *Mu rhythm (de)synchronization and EEG single-trial classification of different motor imagery tasks*, NeuroImage, 31, pp. 153–159

Pfurtscheller, G., Allison, B. Z., Bauernfeind, G., Brunner, C., Solis Escalante, T., Scherer, R., ... & Birbaumer, N. 2010. *The hybrid BCI*. Front in neurosci, 4, 3

Randolph, A.B., 2012. *Not all created equal: individual technology fit of brain–computer interfaces*. In: 45th Hawaii International Conference on System Science HICSS, pp. 572-78.

Sanchez J.C, Mahmoudi B., DiGiovanna J., Principe J.C. 2009. *Exploiting co-adaptation for the design of symbiotic neuroprosthetic assistants*, Neural Networks, 22, 305-315

Sannelli, C., Vidaurre, C., Müller, K.R. & Blankertz, B. 2016. *Ensembles of adaptive spatial filters increase BCI performance: an online evaluation,* Journal of neuroeng, 13, 046003

Satti, A. Guan, C. Coyle, D. & Prasad, G. 2010. *A covariate shift minimisation method to Alleviate non-stationarity Effects for an adaptive brain-computer interface,* Proc. ICPR

Shenoy, P. Krauledat, M. Blankertz, B. Rao, R. & Müller, K.R. 2006. *Towards adaptive classification for BCI*. Journal of NeuroEng, *3*, R13

Shenoy, K.V. and Carmena, J.M., 2014. *Combining decoder design and neural adaptation in brain-machine interfaces*. Neuron, 84(4), pp.665-680.

Schlögl, A.; Vidaurre, C. & Müller, K.-R.2010 *Adaptive methods in bci research-an introductory tutorial*

Song, X.; Yoon, S.-C. & Perera, V. 2013 *Adaptive Common Spatial Pattern for single-trial EEG classification in multi subject BCI*, Proc. IEEE EMBS NER, 411-414

Sun, S. & Zhang, C. 2006. *Adaptive feature extraction for EEG signal classification*, Med and Bio Eng and Comp, Springer, 44, 931-935

Sweller, J., van Merrienboer, J., & Pass, F. 1998. *Cognitive Architecture and Instructional Design Educational* Psychology Review, 10, 251-296

Sykacek, P.; Roberts, S. J. & Stokes, M. 2004. *Adaptive BCI based on variational Bayesian Kalman filtering: an empirical evaluation*, IEEE Transactions on Biomed Eng 51, 719-729

Teillet, S., Lotte, F., N'Kaoua, B., Jeunet, C. 2016. *Towards a Spatial Ability Training to Improve Motor Imagery based Brain-Computer Interfaces (MI-BCIs) Performance: a Pilot Study*, Proc. IEEE SMC

Thomas, K. Guan, C. Lau, C. Prasad, V. & Ang, K. 2011. *Adaptive Tracking of Discriminative Frequency Components in EEG for a Robust Brain-Computer Interface*, Journal of NeuroEng, 8, 1-15

Tan D., Nijholt A. 2010. *Brain-Computer Interaction: Applying our Minds to HCI*. London: Springer-Verlag

Tan, L.F., Dienes, Z., Jansari, A., & Goh, S. Y. 2014. *Effect of mindfulness meditation on brain–computer interface performance*. Consciousness and cognition, 23, 12-21

Tomioka, R.; Hill, J.; Blankertz, B. & Aihara, K. 2006. *Adapting Spatial Filtering Methods for Nonstationary BCIs*, Proc. IBIS, 65-70

Vidaurre, C.; Schlogl, A.; Cabeza, R.; Scherer, R. & Pfurtscheller, G. 2007. *Study of on-line adaptive discriminant analysis for EEG-based brain computer interfaces*. IEEE transactions on biomed eng, 54, 550-556

Vidaurre, C. & Schlogl, A. 2008. *Comparison of adaptive features with linear discriminant classifier for brain computer interfaces*, Proc. EMBC, 173-176

Vidaurre, C.; Kawanabe, M.; Von Bunau, P.; Blankertz, B. & Muller, K. 2011. *Toward unsupervised adaptation of LDA for brain--computer interfaces*, Biomed Eng, IEEE Transactions on, 58, 587-597

Vidaurre, C.; Sannelli, C.; Müller, K.-R. & Blankertz, B. 2011. *Co-adaptive calibration to improve BCI efficiency*. Journal of neuroeng, 8, 025009

Vygotsky, L. S. 1978. *Mind in society: The development of higher psychological processes.* Harvard university press.

Vourvopoulos, A., Ferreira, A. and Bermudez S. 2016. *NeuRow: An Immersive VR Environment for Motor-Imagery Training with the Use of Brain-Computer Interfaces and Vibrotactile Feedback* in Proc PhyCS.

Vuckovic, A, Pineda, J., La Marca, K., Gupta, D. and Guger, C. 2014. *Interaction of BCI with the underlying neurological conditions in patients:pros and cons*, front in neuroeng

Witte, M., Kober, S. E., Ninaus, M., Neuper, C., & Wood, G. 2013. *Control beliefs can predict the ability to up-regulate sensorimotor rhythm during neurofeedback training*.

Wolpaw, J. R. and Wolpaw E. W. 2012. *BCI: Principles and Practice*. Oxford University Press

Woehrle, H.; Krell, M. M.; Straube, S.; Kim, S. K.; Kirchner, E. A. & Kirchner, F. 2015. *An adaptive spatial filter for user-independent single trial detection of event-related potentials*, IEEE Transactions on Biomed Eng, 62, 1696-1705

Zhao, Q.; Zhang, L.; Cichocki, A. & Li, J. 2008. *Incremental Common Spatial Pattern algorithm for BCI*, Proc. IJCNN, 2656 -2659

Zich, C., Debener, S., Chen, L. C., & Kranczioch, C. 2016. *FV 10. Motor imagery supported by neurofeedback: Age-related changes in EEG and fNIRS lateralization patterns*. Clinical Neurophysiology, 127(9), e215.